\documentclass[journal=apcach,manuscript= perspective]{achemso}

\usepackage{bm}

\usepackage{chemformula} 
\usepackage[T1]{fontenc} 
\usepackage{multimedia}

\newcommand{\limtinf}{\underset{t\rightarrow\infty}{\lim}}
\newcommand{\mygam}{\mathbf{\Gamma}^{'}}

\author{Debra J. Searles}
\email{d.bernhardt@uq.edu.au}
\affiliation{Australian Institute for Bioengineering and Nanotechnology, The University of Queensland, Brisbane, QLD, 4072, Australia}
\alsoaffiliation{School of Chemistry and Molecular Biosciences, The University of Queensland, Brisbane, QLD, 4072, Australia}

\author{Stephen Sanderson}
\affiliation{Australian Institute for Bioengineering and Nanotechnology, The University of Queensland, Brisbane, QLD, 4072, Australia}

\title[Fluctuation Theorems and Distribution Functions for Polar Molecules]
{Fluctuation theorems and distribution functions for polar molecules in an electric field}

\abbreviations{}

\begin{document}


\begin{abstract}
  In this perspective we consider how modern statistical mechanics and response theory can be applied to understand the response of polar molecules to an applied electric field and the fluctuations in these systems. Results that are consistent with thermodynamics and physical expectations are derived, as well as  a new fluctuation relation that is tested in molecular simulations. 
  It is demonstrated that a deterministic approach leads to distribution functions that continually evolve, even when system properties have relaxed, which has implications for treatment of nonequilibrium steady states. 
\end{abstract}

\section{Introduction}
The development of fluctuation theorems has led to the understanding of how the microscopic dynamics of a system leads to the macroscopic response to a change in conditions or application of a field or mechanical driving force.\cite{evans2016fundamentals,Sevick2008603}  There are several different considerations to be made when developing these results, including whether the dynamics is deterministic or not, and any explicit time-dependence of the equations of motion. The fluctuation relationships obtained have provided a statistical mechanical derivation of thermodynamic principles such as Le Chatelier's principle,\cite{evans2001,dasmeh2009,dasmeh2011} the Second Law of Thermodynamics,\cite{Evans1993,Evans1994,Evans1996,evans2012mathematical,Jarzynski2011,Carberry2004140601,Wang2002PRL} Clausius's Theorem,\cite{Evans2011,Evans2016aderivation} Gibbs' Equation,\cite{Evans2016aderivation} Fourier's Law\cite{Evans2010Fourier,evans2012mathematical,Onuki2019} and the Zeroth Law of Thermodynamics.\cite{evans2016fundamentals}   Furthermore, they provide a relationship for the probability that for a small system, monitored for short times, there will be a change in the system contrary to that predicted by the relevant thermodynamic principle.\cite{petersen2016dissipation,petersen2016mechanism,evans2016fundamentals,Wang2002PRL} The references provided above are mostly to derivations based on deterministic dynamics, which will be the focus of this perspective article.

Le Chatelier's principle considers the response of a system at equilibrium to a change that drives it to another equilibrium state, predicting that initial deviations from equilibrium will spontaneously decay rather than grow. The link between fluctuation theorems and Le Chatelier's principle has been directly demonstrated for systems subject to a temperature change.\cite{dasmeh2009,dasmeh2011} In a similar way, a Hamiltonian system with an initial perturbation, evident as a non-uniform density profile, was shown to relax to a uniform density.\cite{evans2001}  

In related work on deterministic dynamical systems, it has been shown how the form of the equilibrium distribution function can be obtained from the dynamics of the system, if it exists.\cite{petersen2022} The Liouville equation (or phase space continuity equation) can be used to show how a phase space distribution function will evolve with time according to given dynamics, and the existence of a normalisable, time-independent phase space distribution for all fixed phase points implies that an equilibrium distribution exists. One of the systems considered in ref. \citenum{petersen2022} was a system of charged particles subject to an electric field. It was shown that only in the case of a system with bounds in the direction of the field could an equilibrium distribution be obtained. However, in that manuscript it was assumed that the charged particles were not bonded to each other.  Here we extend that case to one where the charges are connected to form charge-neutral molecules overall, as in the case of polar molecules. A trivial extension of that work leads to an equilibrium distribution function for periodic systems subject to field, and the distribution function for the thermostatted system is a Boltzmann distribution including the potential energy due to the field, as expected.

We then consider the fluctuation theorem for a system of polar molecules to which a field is applied at time-zero, and prove that the molecules will orient so that the dipoles align in the direction of the field, on average, as expected by the second law of thermodynamics. Furthermore, we show the likelihood that for small systems monitored over short times, reorientation in a direction different to that expected thermodynamically will occur. In a similar manner, we show that the Crooks Fluctuation Theorem \cite{Crooks1999} and the Jarzynski equality \cite{Jarzynski2011} can be used to determine the free energy difference between the systems with and without alignment.

Finally, the Liouville equation provides a relationship for the time-evolution of the distribution function, and in this case (application of a field) we know what it should relax to. However, it is also known that for thermostatted, field-driven processes the distribution function continually changes. We therefore examine this asymptotic relationship to see how this apparent paradox can be reconciled. 

\section{Results and discussion}

\subsection{The distribution function for polar molecules subject to a field} \label{sec:eq_df}

In this work we consider a system of charge-neutral molecules, some of which are polar.  For simplicity, we consider a non-polar solvent containing polar diatomic molecules that are modelled by placing a point charge at each of the atomic sites, although the theory is readily extended to other cases provided each molecule is charge-neutral. The equations of motion for each atom in this system, subject to a an external electric field are:

\begin{eqnarray} \label{eq:eqnmot}
{\mathbf{\dot q}_{i}} &=& {\mathbf{p}_{i}/m_{i}} \nonumber \\
{\mathbf{\dot p}_{i}} &=& {\mathbf{F}_{i}} + {\mathbf{i}}{F_e}{z_{i}} - S_i\alpha {{\mathbf{p}}_{i}} \nonumber \\
\dot \alpha  &=&  \frac{1}{Q}\left(\sum_{i}^{N}\frac{S_i{\bf p}_{i}^2}{m_{i}}-3N_{th}k_BT\right).
\end{eqnarray}
Here $\mathbf{q}_{i}$ is the position of atom $i$, $\mathbf{p}_{i}$ is its momentum, $m_{i}$ is its mass and $\mathbf{F}_{i}$ is the interatomic force on each atom (which could include Coulombic forces). An electric field, $F_e$ is applied in the x-direction with $\mathbf{i}$ being the unit-vector in the x-direction.  Since application of a field will do work on the system, a thermostat is used to prevent the system from heating. A Nos{\'e}-Hoover thermostat is selected in this work, with a switch, $S_i$=0 or 1 indicating which of the the molecules are thermostatted. This would generate the Nos{\'e}-Hoover canonical ensemble in the field-free case.\cite{petersen2022}  In our system, a constant field is applied at $t=0$, whereas the Nos{\'e}-Hoover thermostat is applied at all times. The total number of atoms is given by $N$, and the number of thermostatted atoms is $N_{th}$. The target temperature, $T$, and the effective mass of the thermostat, $Q$, are constant parameters. The temperature is related to the ensemble average of the kinetic energy of the thermostatted particles, $3N_{th}k_BT=\langle 2K_{th}\rangle=\langle \sum_{i}^{N}({S_i{\bf p}_{i}^2}/{m_{i}})\rangle$ where $k_B$ is Boltzmann's constant.

Rather than the general case considered previously,\cite{Petersen20226383} here we assume that the charge-sites are on molecules that have net-zero charge. However, we apply the same procedure as in Petersen et al.\cite{petersen2022} to calculate the form of the equilibrium distribution given the equations of motion that describe the system. That is, we assume a general form of the distribution function that includes an arbitrary phase function ($G(\bm{\Gamma}))$, where $\bm{\Gamma}  = ({\mathbf{q}},{\mathbf{p}})$) and determine if there is a phase function $G(\bm{\Gamma})$ that ensures that the distribution function at all $\bm{\Gamma}$ is fixed, never negative and can be normalised.  

Similarly to Petersen et al.\cite{petersen2022}, we choose a general form of the distribution function to be, 

\begin{eqnarray}
f_{F_e}(\bm{\Gamma},\alpha) = \frac{{\exp \left( { - \beta \left(H_0({\bm{\Gamma} }) + \frac{1}{2}Q\alpha^2  + G({\bm{\Gamma}} ) \right)} \right)\delta (\mathbf{P}(\bm{\Gamma} ))}}{{\int  \exp \left( { - \beta \left(H_0({\bm{\Gamma} }) + \frac{1}{2}Q{\alpha}^2  + G({\bm{\Gamma} }) \right)} \right)}\delta (\mathbf{P}(\bm{\Gamma} ))d \bm{\Gamma}d\alpha}.
\label{eq:f_general}
\end{eqnarray}
which is a perturbation of the the field-free distribution function considered. Here, $\beta={1}/({k_B T})$, $H_0(\bm{\Gamma})$ is the field-free internal energy, and $\delta(\mathbf{P}(\bm{\Gamma} ))$ indicates that the total momentum of the system is fixed, consistent with the molecular dynamics ensemble.\cite{} In order for the conditions for the existence of an equilibrium distribution function to be met, it is necessary that,\cite{petersen2022}
\begin{equation}
\begin{split}
\beta\left[H_0({\bm{\Gamma}(t) }) + \frac{1}{2}Q\alpha(t)^2  + G({\bm{\Gamma}(t)})\right]-\int_0^t \Lambda(\bm{\Gamma}(s),\alpha(s))ds\\
=\beta \left[H_0({\bm{\Gamma}(0) }) + \frac{1}{2}Q\alpha(0)^2  + G({\bm{\Gamma}(0)})\right],  
\end{split}
\end{equation}
where $\Lambda$ is the phase space expansion rate, or equivalently,
\begin{equation}
\beta(\dot H_0 + Q\alpha \dot \alpha + \dot G)-\Lambda
=0.  \label{eq:condition}
\end{equation}
The phase space expansion rate for this dynamics is,
\begin{equation}
\Lambda  = \frac{\partial }{{\partial \bm{\Gamma} }} 
\cdot \dot {\bm{\Gamma}} + \frac{\partial }{{\partial \alpha }} 
 \dot {\alpha} =  - 3N_{th}\alpha,	\label{eq:colL}
\end{equation}
and the change in the internal energy with time is given by the sum of the changes in the kinetic and potential energies,
\begin{equation}
\dot H_0 = \frac{{\mathbf{p}} \cdot {\mathbf{\dot p}}}{m} - {\mathbf{F}}\cdot {\mathbf{\dot q}} = F_e \sum\limits_{i}^{N} {z_{i}{v_{x,i}}}  - \sum_{i}^{N}\frac{S_i{\bf p}_{i}^2}{m_{i}}\alpha.  \label{eq:colH}
\end{equation}
If the equations of motion generate an equilibrium distribution, we will be able to find a $G$ which has no explicit time-dependence and the exponent in \ref{eq:f_general} is bounded above for all ${\bm{\Gamma}}$.\cite{petersen2022} Substitution of (\ref{eq:colL}) and (\ref{eq:colH}) and (\ref{eq:eqnmot}) in (\ref{eq:condition}) gives
\begin{equation}
    \dot G(\bm{\Gamma}) = -F_e \sum\limits_{i}^{N} {z_{i}{v_{x,i}}}.
\end{equation}
A solution is,
\begin{equation}
    G(\bm{\Gamma}) = -F_e \sum\limits_{i}^{N} z_{i}{q_{x,i}},
\end{equation}
which has no explicit time-dependence, as required. However, in an unbounded or periodic system there is no bound on $q_{x,i}$, so it might seem that $G(\bm{\Gamma})$ will also be unbounded and an equilibrium distribution function would not exist.\cite{petersen2022}. Yet, this is not the case for charge-neutral polar molecules. The dipole moment in the x-direction of molecule $j$ is given by $\mu_{x,j}=\sum_k^{N_{at}}z_kq_{x,k}$ where $N_{at}$ is the number of atoms in molecule $j$ and then, 
\begin{equation}
    G(\bm{\Gamma}) =  -F_e \sum\limits_{j}^{N_{mol}} {\mu_{x,j}(\bm{\Gamma})},
\end{equation}
where we now sum over all $N_{mol}$ molecules. In the limit of long bond length, the harmonic contribution to the potential energy will grow more quickly than the dipole moment, so $H_0({\bm{\Gamma} }) + \frac{1}{2}Q\alpha^2  + G({\bm{\Gamma}})$ is bounded below for any finite field and the phase space distribution function will become, 
\begin{equation} \label{eq:canonical_field}
f_{F_e}(\bm{\Gamma},\alpha ,0) = \frac{{\exp ( { - \beta (H_0(\bm{\Gamma} )+ \frac{1}{2}Q\alpha^2 - {F_e}\sum\limits_{j = 1}^{N_{mol}} \mu_{x,j}(\bm{\Gamma} ) )} )}\delta (\bm{P(\bm{\Gamma})} ) }{{\int {\exp ( { - \beta (H_0(\bm{\Gamma} ) + \frac{1}{2}Q\alpha^{2}  - {F_e}\sum\limits_{j = 1}^{N_{mol}} \mu_{x,j}(\bm{\Gamma} ) )} )\delta (\bm{P}(\bm{\Gamma} ) ) d\bm{\Gamma} }d\alpha }}.
\end{equation}
The dependence on $\alpha$ can be integrated out if desired.  This is not a surprising result: it is just the Nos{\'e}-Hoover thermostat canonical distribution where the internal energy includes that due to the field. However, it is instructive to see how the derivation differs to that of a system of ions in a bounded cell, a periodic system or an unbounded system. 

\subsection{Fluctuation theorems for the response of polar molecules to a field } \label{sec:FT_dG}
The fluctuation theorem gives an expression for the probability ratio that a function known as the \textit{dissipation function} takes on a positive or negative values of the same magnitude.  The dissipation function is uniquely defined in a deterministic system with known initial distribution and dynamics.  For the case of a solution of polar molecules described above, we can consider the response of a field-free system that is at equilibrium at $t=0$ to the application of a field.  Therefore, the initial distribution function is the Nos{\'e}-Hoover thermostat canonical distribution,
\begin{equation} \label{eq:canonical}
f(\bm{\Gamma},\alpha) = \frac{{\exp ( { - \beta (H(\bm{\Gamma} )+ \frac{1}{2}Q\alpha^2 )} )}\delta (\bm{P(\bm{\Gamma})} ) }{{\int {\exp ( { - \beta (H(\bm{\Gamma} ) + \frac{1}{2}Q\alpha^{2}   )} )\delta (\bm{P}(\bm{\Gamma} ) ) d\bm{\Gamma} } d\alpha }},
\end{equation}
and the dynamics are given by equation (\ref{eq:eqnmot}). The extended phase space is represented by $\bm{\Gamma^{'}}\equiv(\bm{\Gamma},\alpha)$ below. 

To determine the dissipation function, one compares the ratio of probabilities of observing phase points in phase volumes about a forward trajectory ($d\bm{\Gamma^{'}} \to d\bm{\Gamma}^{'}(t))$) and its time-reversed conjugate ($d\bm{\Gamma}^{'*} \to d\bm{\Gamma}^{'*}(t)$), (\textit{i.e} the trajectory that would be obtained if time ran backwards from the end-point of the initial trajectory).  For time-reversible dynamics the time-reversed conjugate trajectory can be obtained by carrying out a time-reversal mapping of the end point of the original trajectory and then evolving in time for the same period. For our equations of motion the time-reversal mapping is $(\mathbf{q},\mathbf{p},\alpha) \to (\mathbf{q},-\mathbf{p},-\alpha)$.  Since the dynamics is deterministic, the probability of observing an initial phase volume is equal to the probability of observing the trajectories originating from that phase volume. That is,\cite{Searles20003503,evans2016fundamentals}
\begin{equation} \label{eq:dissfn}
	{\Omega_t(\bm{\Gamma}^{'})}\equiv  {\int_0^t\Omega(\bm{\Gamma}^{'}(s))ds}\equiv \ln\frac{p(d\bm{\Gamma} \to d\bm{\Gamma}^{'}(t))}{p(d\bm{\Gamma}^{'*} \to d\bm{\Gamma}^{'*}(t))}=\ln \frac{p(d\bm{\Gamma}^{'})}{p(d\bm{\Gamma}^{'*})}= \ln\frac{f(\bm{\Gamma^{'}};0)d\bm{\Gamma^{'}}}{f(\bm{\Gamma}^{'*};0)d\bm{\Gamma}^{'*}}
\end{equation}
where $\Omega_t(\bm{\Gamma}^{'})$ is the time-integral of the dissipation function, $\Omega(\bm{\Gamma}^{'})$, and we use the notation $f(\bm{\Gamma^{'}};t)$ to denote the phase space density about a point $\bm{\Gamma^{'}}$, according to the  distribution function at time $t$.
For the combination of dynamics (\ref{eq:eqnmot}) and initial distribution function (\ref{eq:canonical}), equation (\ref{eq:dissfn}) gives the integral of the dissipation function,
\begin{equation} \label{eq:dipole_dissfn}
	\Omega_t= \beta F_e \sum_j^{N_{mol}} (\mu_{x,j}(t)-\mu_{x,j}(0))=\beta F_e N_{mol} \overline{\Delta_t \mu}_x,
\end{equation}
where $\overline{\Delta_t \mu}_x$ is the average change in the x-dipole moment per molecule over the period $t$. The dissipation function can then be obtained, $\Omega=\beta F_e \sum_j^{N_{mol}} \dot{\mu}_{x,j} =\beta F_e \sum_i^Nz_iv_{x,i}$.

The fluctuation theorem predicts,
\begin{equation} \label{eq:FT_dipole}
	\frac {p(\overline{\Delta_t \mu}_x=A)}{p(\overline{\Delta_t \mu}_x=-A)}= e^{\beta F_e N_{mol} A}
\end{equation}
where the notation $p(X=A)$ implies the probability that the property $X$ takes on a value in the range $A\pm dA$. From this, 
\begin{equation} \label{eq:2nd_law}
	\beta F_e N_{mol}\langle \overline{\Delta_t \mu}_x \rangle \ge 0.
\end{equation}
That is, (\ref{eq:2nd_law}) states the change in the x-dipole moment of the entire system on application of a positive field will be positive on average and (\ref{eq:FT_dipole}) further indicates that the likelihood of observing negative changes will decrease with increasing field, number of polar molecules and with decreasing temperature.  

The result given by (\ref{eq:FT_dipole}) is often referred to as the \textit{transient fluctuation theorem} and considers the likelihood of observing behaviour in response to application of a field. In this case, the steady state fluctuation theorem becomes trivial since $\limtinf \frac{1}{t}\overline{\Delta_t \mu}_x =0 $, \textit{i.e.} the dipole moment will fluctuate around a steady value in the long time limit.  

The Crooks fluctuation theorem differs in that it compares the probability of observing trajectories sampled from a field-free canonical ensemble and the ensemble with a field applied. It determines the free energy difference between these systems in terms of the probability that the work takes on different values, where the work carried out over a period $t$, $w_t$, is defined as:
\begin{equation} \label{eq:work}
	\beta {w_t(\bm{\Gamma}^{'})}\equiv  \ln\frac{p_{F_e=0}(d\bm{\Gamma} \to d\bm{\Gamma}^{'}(t))}{p_{F_e}(d\bm{\Gamma}^{'*} \to d\bm{\Gamma}^{'*}(t))}=\ln \frac{p_{F_e=0}(d\bm{\Gamma}^{'})}{p_{F_e}(d\bm{\Gamma}^{'*})}= \ln\frac{f_{F_e=0}(\bm{\Gamma^{'}};0)d\bm{\Gamma^{'}}}{f_{F_e}(\bm{\Gamma}^{'*};0)d\bm{\Gamma}^{'*}}
\end{equation}
For the combination of dynamics (\ref{eq:eqnmot}), initial distribution function (\ref{eq:canonical}) and final distribution function (\ref{eq:canonical_field}), the work is,
\begin{equation} \label{eq:dipole_work}
	w_t= -F_e \sum_j^{N_{mol}} \mu_{x,j}(0)
\end{equation}
This indicates that the work over a trajectory of any length will be the same.  This is because the work is only done at $t=0$ with application of the field and is given by the initial dipole moment of the system.  If the field is applied gradually, the work will be given by the integral over that period. The Crooks fluctuation theorem is,
\begin{equation} \label{eq:CrooksFT}
	\frac {p_{F_e=0}(w_t^F=A)}{p_{F_e}(w_t^R=-A)}= e^{\beta(A-\Delta F)},
\end{equation}
where $\Delta F$ is the change in Helmholtz free energy, and for this case where the field is applied instantaneously, it becomes

\begin{equation} \label{eq:CrooksFT2}
	\frac {p_{F_e=0}(\overline {\mu_x}(0)=A)}{p_{F_e}(\overline {\mu_x}(0)=-A)}= e^{-\beta(F_e N_{mol} A+\Delta F)}
\end{equation}
and the corresponding Jarzynski equality is,
\begin{equation} \label{eq:JE}
	\Delta F = -k_B T \ln \langle e^{\beta F_e N_{mol} \overline {\mu_x}(0)}\rangle_{F_e=0}
\end{equation}
with the Jarzynski equality clearly giving an expression for the free energy that is consistent with that obtained directly from the partition function of the two equilibrium distribution functions (eqns (\ref{eq:canonical}) and (\ref{eq:canonical_field})).

\subsection{Relaxation to equilibrium} \label{sec:relax}
The Liouville equation shows how a distribution function will evolve in time.  It can be written, 

\begin{equation} \label{eq:Liouville}
	f(\bm{\Gamma};t) = e^{ \int_0^t \Omega(\bm{\Gamma}(-s))ds}f(\bm{\Gamma};0) = e ^{ \int_0^t \Omega(\bm{\Gamma}^*(s))ds}f(\bm{\Gamma};0)
\end{equation}
where the second equality uses the relationship between the trajectory and its time-reversed conjugate.  The case we consider in this manuscript provides an interesting application because an initial field-free equilibrium state evolves to another equilibrium state that has a field applied. We therefore know the asymptotic expression for the distribution function that the system is relaxing to. It is therefore of interest to see how this convergence occurs.

Using the dissipation function determined above, we can write:
\begin{equation} \label{eq:Liouville-charge}
	f(\bm{\Gamma}^{'};t) = e^{ \beta F_e (\mu_{x,tot}(\bm{\Gamma}^{'})-\mu_{x,tot}(\bm{\Gamma}^{'}(-t)))}f(\bm{\bm{\Gamma}^{'}};0)=e^{ -\beta F_e \mu_{x,tot}(\bm{\Gamma}^{'}(-t))}f_{F_e}(\bm{\bm{\Gamma}^{'}};0)\frac{Z_0}{Z_{F_e}},
\end{equation}
where $Z_0$ and $Z_{F_e}$ are the partition functions of the Nos{\'e}-Hoover canonical distributions without and with the field applied, respectively. This indicates that the phase space distribution function continually evolves and will not relax to the the field-dependent Nos{\'e}-Hoover canonical distribution, $f_{F_e}(\bm{\bm{\Gamma}^{'}};0)$, even at long times. This seems at odds with the fact that we expect the system to relax to a steady (equilibrium) state where properties will no longer change.  

We can resolve this apparent contradiction. The phase space density is not an observable, and what we are really interested in from a physical point of view is properties that can be expressed as ensemble averages of phase variables, $\langle B(t)\rangle$.  In our chosen system, we would expect that the properties of the field-free equilibrium system will evolve to those of the system with the field on.  Therefore we consider the evolution of the phase variables with time,
\begin{equation}
	\begin{split}
	\langle B(t)\rangle&=\int B(\mygam) f(\mygam,t) d\mygam\\
	&=\int B(\mygam) e^{-\beta F_e \mu_{x,tot}(\mygam(-t))} f_{F_e}(\mygam,0)d\mygam \frac{Z_{F_e}}{Z_0}\\
	&=\langle B(\mygam) e^{-\beta F_e \mu_{x,tot}(\mygam(-t))} ) \rangle_{F_e} \frac{Z_{F_e}}{Z_0}.
	\end{split}
\end{equation}
Then in the long time limit and if decorrelation occurs we can split the ensemble average in two,
\begin{equation}
	\begin{split}
		\limtinf \langle B(t)\rangle&=\langle B(\mygam) \rangle_{F_e} \langle e^{-\beta F_e \mu_{x,tot}(\mygam(-t))} ) \rangle_{F_e} \frac{Z_{F_e}}{Z_0}.
	\end{split}
\end{equation}
Consider the average $\langle e^{-\beta F_e \mu_{x,tot}(\mygam(-t))} \rangle_{F_e} \equiv  \langle e^{-\beta F_e \mu_{x,tot}(\bm{\Gamma}^{'*}(t))} \rangle_{F_e}$. Since the dynamics is field-driven and the ensemble average is over the equilibrium ensemble sampled by that dynamics, the ensemble average is time-independent and is equal to $\langle e^{-\beta F_e \mu_{x,tot}(\mygam)} \rangle_{F_e}$, so

\begin{equation}
\label{eq:phase_var_asymp}
	\begin{split}
		\limtinf \langle B(t)\rangle&=\langle B(\mygam) \rangle_{F_e} \langle e^{-\beta F_e \mu_{x,tot}(\mygam)} ) \rangle_{F_e} \frac{Z_{F_e}}{Z_0}\\
		&=\langle B(\mygam) \rangle_{F_e} \int  e^{-\beta F_e\mu_{x,tot}(\mygam)} \frac{ e^{-\beta(H_0(\mygam)-F_e\mu_{x,tot}(\mygam)}}{Z_{F_e}}   d \mygam \frac{Z_{F_e}}{Z_0}\\
		&=\langle B \rangle_{F_e}
	\end{split}
\end{equation}

Therefore, although the phase space distribution function is constantly evolving as shown by equation (\ref{eq:Liouville-charge}), if correlations in properties decay (the system is mixing), then the properties can be shown to relax to those of the new equilibrium, with the field on. Decay of correlations is typical for chaotic systems and thus for most systems of many interacting molecules. However, it also points out that if correlations do not decay, the properties will not relax.

We can extend this so that rather than the phase-space distribution function, we consider a coarse-grained measure by integrating over a small phase volume element about $\bm{\Gamma}^{'}$,

\begin{equation}
	\begin{split}
		\int_{\delta V} f(\mygam;t) d\mygam &=	\int W_{\delta V}(\mygam) f(\mygam;t) d\mygam \\
			&=\langle W_{\delta V}(t) \rangle \stackrel{\limtinf}{=} \langle W_{\delta V} \rangle_{F_e}\\
			&= 	\int_{\delta V} f_{F_e}(\mygam;0) d\mygam,
	\end{split}
\end{equation}
where $W_{\delta V}(\mygam)$ is a window function that selects a region of points in $\delta V$ about $\mygam$ and we use the same assumptions on decay of correlations that we used to obtain (\ref{eq:phase_var_asymp}).  This indicates that for chaotic systems, in the long time limit the coarse-grained distribution function converges as expected.

\subsection{Numerical results} \label{sec:numerical}
To illustrate some of the results above in a practical way, we carry out molecular dynamics simulations of dipolar molecules in a solvent. We use Lennard-Jones reduced units throughout this section.  Three-dimensional periodic systems of 108 molecules at a number density of 0.8 and temperature 1.0 were modelled with 1, 10 or 50 polar diatomic molecules, AB, in a solvent. The solvent molecules had no charge, a mass of 1 and Lennard-Jones parameters of $\sigma$=1 and $\epsilon$=1. Atom A of the polar molecule had a mass 0.8, charge of +10, $\sigma$=1 and $\epsilon$=1 whereas atom B of the polar molecule had  mass 0.7, charge of -10, $\sigma$=1 and $\epsilon$=1. A Lennard-Jones interaction cutoff of 2.5 was used in all cases. The bond of each AB molecule was modelled using a harmonic potential with a force constant of 200 and a equilibrium bond length of 0.2. The total momentum of the system was initially set to zero.  The Nos{\'e}-Hoover thermostatted equations of motion (\ref{eq:eqnmot}) were integrated with a timestep of 0.001 and a damping factor, $Q=0.025$.  The LAMMPS \cite{LAMMPS} simulation package was used for all molecular dynamics simulations. 

Initially the equilibrium distributions of the systems were sampled with no field applied. The system was equilibrated for 10,000 timesteps, and then 500 points from the equilibrium distribution were sampled by saving the phase point every 1000 timesteps. A field of 0.1 or 2.0 was applied and the evolution of the dipole moment was monitored.

\begin{figure}
    \centering
    \includegraphics{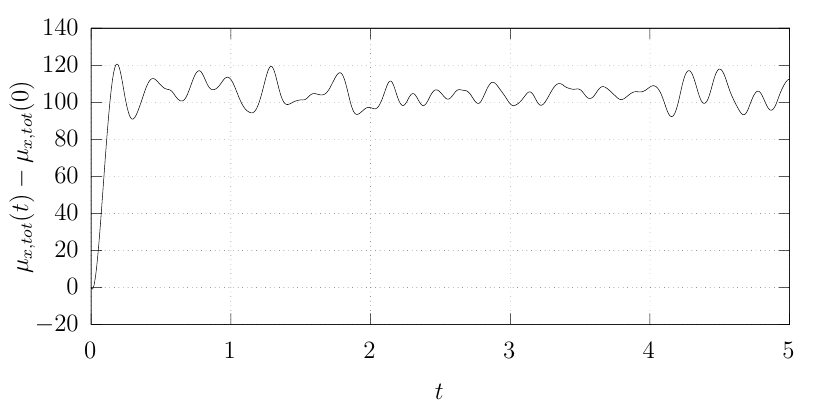}
  \caption{The change in the dipole moment for a trajectory that is sampled from field-free equilibrium and responds to the application of a field.  The system consists of 50 polar molecules in 58 non-polar solvent molecules at a temperature of 1.0 and molecule density of 0.8, and the field is 2.0 units.}
  \label{fig:response}
\end{figure}

Figure \ref{fig:response} shows the change in the total dipole moment for a single trajectory of the system of 50 polar molecules over 5 time units (5000 timesteps) after application of a field of 2.0 units. Clearly, there is an immediate increase in the dipole moment, which then oscillates about a constant value after the field has been applied for $\approx 2$ time units. If all polar molecules were aligned with the field and at their equilibrium geometry, a dipole moment of 100 would be obtained for this system. Changes in the dipole moment that are greater than 100 occur due to the vibrations of the molecules and because the initial dipole moment is non-zero. Oscillations occur due to vibrational and rotational motion of the molecules. In the supporting material, animations of the response of the molecules to the field of 2.0 are provided. In one case all the molecules are shown and in the other only the polar molecules are shown.  At the start of the simulation, the molecules are randomly oriented, and by the end a preferred orientation is evident. 

\begin{figure}
    \centering
    \includegraphics{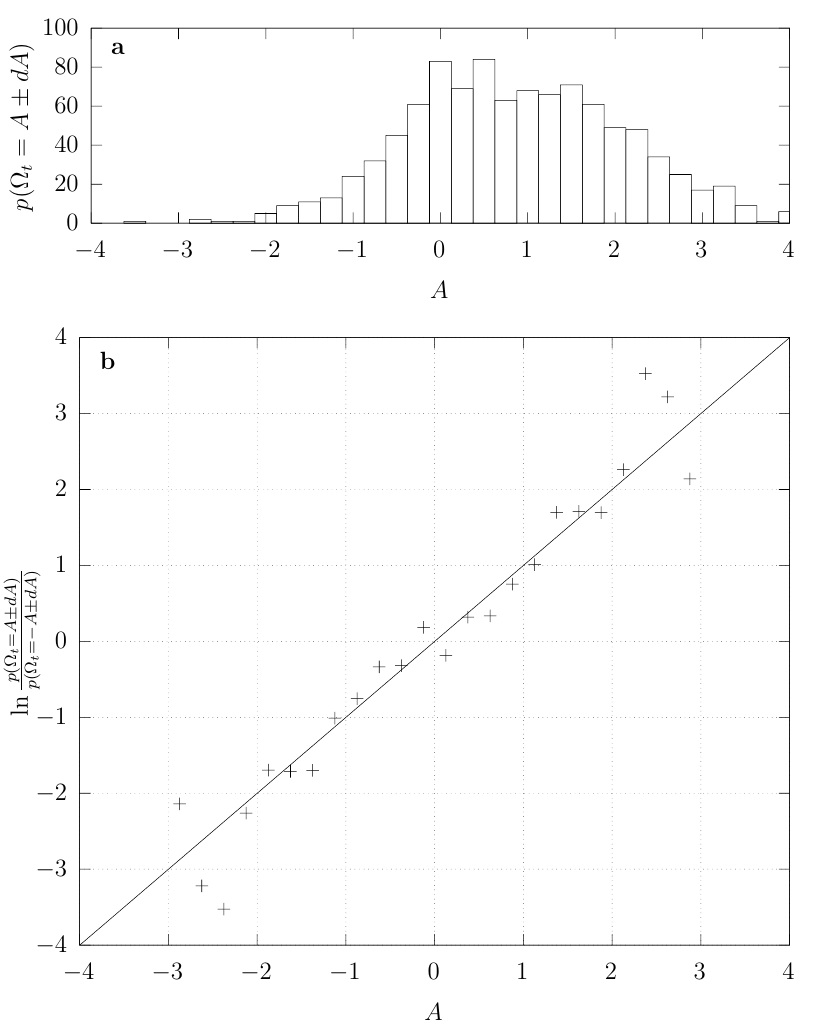}
  \caption{(a) A histogram of the time-integral of the dissipation ($\Omega_t=\beta F_e [\mu_{x,tot}(t)- mu_{x,tot}(0)]$) for 984 trajectories that are sampled from field-free equilibrium and respond to the application of a field.  The system consists of 10 polar molecules in 98 solvent molecules at a temperature of 1.0, molecule density of 0.8 and the field is 0.2 units.  The dissipation is proportional to the change in dipole moment and is monitored over 0.2 units of time. (b) A test of the fluctuation theorem for this system. The theoretical prediction is shown by the straight line of unit slope.}
  \label{fig:FT}
\end{figure}

A set of 984 nonequilibrium trajectories sampled from the equilibrium distribution where then simulated. Fig.\ref{eq:FT_dipole}a shows the distribution of the time-integral of the dissipation function ($\Omega_t=\beta F_e [\mu_{x,tot}(t)- mu_{x,tot}(0)]$) for a system of 10 polar molecules, monitored over 0.2 time units (200 timesteps), with a field of 0.2 units. The change in the dipole moment in the x-direction, which is proportional to the dissipation function, is expected to be positive on average and this is evident from the histogram. As the field increases, the temperature decreases, or the number of particles increases, the histogram will shift to higher values of the dissipation, and negative dissipation (decrease in the dipole moment) will be less common. The system shown in Fig.\ref{fig:FT} was selected so that trajectories with negative dissipation could be readily observed and then the fluctuation theorem (\ref{eq:FT_dipole}) could be tested as shown in Fig.\ref{fig:FT}b.  The data is consistent with the prediction which is given by the straight line.

\subsection{Conclusions} 
\label{sec:conclusions}
In this perspective, we have considered how statistical mechanical approaches for deterministic systems that have been developed over the past few decades can be applied to polar molecules that are subject to an electric field. The equilibrium distribution function for the system and fluctuation relations are derived with the results obtained being consistent with thermodynamics expectations and also providing relationships for fluctuations in small systems monitored for short times. We highlight the difference between systems with and without period boundaries; and between systems with ions or neutral molecules even though they can all be modelled using the same equations of motion. The results for the neutral, polar molecules in periodic boundaries are tested in molecular simulations to provide an illustration of the results, and also to show that they can be applied in practice.

Further to this, it is shown that when a system changes from one equilibrium state to another, the distribution function retains memory of the initial state even if the system properties relax to those of the new state. The contradiction that this poses is rationalised by the fact that measurable properties do not require knowledge of the details of the distribution function, but just a coarse-grained measure.

This work serves as a pedagogic illustration of contemporary statistical mechanics and response theory, and also provides new results for the polar molecules subject to a field. In addition, it provides a foundation for further study of systems that reach nonequilibrium steady states where the distribution functions are unknown and might not exist.

\begin{acknowledgement}

The authors thank the Australian Research Council for its support for this project through the Discovery program (FL190100080). They also acknowledge support from the Queensland Cyber Infrastructure Foundation and the University of Queensland Research Computing Centre. The authors thank the participants of the Workshop and School on Theory and Simulation of Nonequilibrium Fluids held at The University of Queensland in September 2023 for their feedback and comments on an incomplete version of this work.  

\end{acknowledgement}

\begin{suppinfo}

The following  are available as ancillary files.
\begin{itemize}
  \item Movie\_all.mpg: An animation showing the response of the simulated system to application of a field. There are 50 polar molecules in a solvent and a field of strength 2.0 units is applied. The solvent molecules are red, the positive atoms of the polar diatomic molecules are green and the negative atoms are blue. At the start of the simulation, the molecules are randomly oriented, and by the end a preferred orientation is evident.   
  \item Movie\_polar.mpg: Same as Movie\_all.mpg, but the solvent is not shown to make the dynamics of the polar molecules more evident. 
\end{itemize}

\

\end{suppinfo}

\bibliography{Polar_FT}

\providecommand{\latin}[1]{#1}
\makeatletter
\providecommand{\doi}
  {\begingroup\let\do\@makeother\dospecials
  \catcode`\{=1 \catcode`\}=2 \doi@aux}
\providecommand{\doi@aux}[1]{\endgroup\texttt{#1}}
\makeatother
\providecommand*\mcitethebibliography{\thebibliography}
\csname @ifundefined\endcsname{endmcitethebibliography}
  {\let\endmcitethebibliography\endthebibliography}{}
\begin{mcitethebibliography}{24}
\providecommand*\natexlab[1]{#1}
\providecommand*\mciteSetBstSublistMode[1]{}
\providecommand*\mciteSetBstMaxWidthForm[2]{}
\providecommand*\mciteBstWouldAddEndPuncttrue
  {\def\EndOfBibitem{\unskip.}}
\providecommand*\mciteBstWouldAddEndPunctfalse
  {\let\EndOfBibitem\relax}
\providecommand*\mciteSetBstMidEndSepPunct[3]{}
\providecommand*\mciteSetBstSublistLabelBeginEnd[3]{}
\providecommand*\EndOfBibitem{}
\mciteSetBstSublistMode{f}
\mciteSetBstMaxWidthForm{subitem}{(\alph{mcitesubitemcount})}
\mciteSetBstSublistLabelBeginEnd
  {\mcitemaxwidthsubitemform\space}
  {\relax}
  {\relax}

\bibitem[Evans \latin{et~al.}(2016)Evans, Searles, and
  Williams]{evans2016fundamentals}
Evans,~D.~J.; Searles,~D.~J.; Williams,~S.~R. \emph{Fundamentals of classical
  statistical thermodynamics}; Wiley: Weinheim, 2016\relax
\mciteBstWouldAddEndPuncttrue
\mciteSetBstMidEndSepPunct{\mcitedefaultmidpunct}
{\mcitedefaultendpunct}{\mcitedefaultseppunct}\relax
\EndOfBibitem
\bibitem[Sevick \latin{et~al.}(2008)Sevick, Prabhakar, Williams, and
  Searles]{Sevick2008603}
Sevick,~E.; Prabhakar,~R.; Williams,~S.; Searles,~D. Fluctuation theorems.
  \emph{Annual Review of Physical Chemistry} \textbf{2008}, \emph{59},
  603--633\relax
\mciteBstWouldAddEndPuncttrue
\mciteSetBstMidEndSepPunct{\mcitedefaultmidpunct}
{\mcitedefaultendpunct}{\mcitedefaultseppunct}\relax
\EndOfBibitem
\bibitem[Evans \latin{et~al.}(2001)Evans, Searles, and Mittag]{evans2001}
Evans,~D.~J.; Searles,~D.~J.; Mittag,~E. Fluctuation theorem for Hamiltonian
  Systems: Le Chatelier's principle. \emph{Physical Review E} \textbf{2001},
  \emph{63}, 051105\relax
\mciteBstWouldAddEndPuncttrue
\mciteSetBstMidEndSepPunct{\mcitedefaultmidpunct}
{\mcitedefaultendpunct}{\mcitedefaultseppunct}\relax
\EndOfBibitem
\bibitem[Dasmeh \latin{et~al.}(2009)Dasmeh, Searles, Ajloo, Evans, and
  Williams]{dasmeh2009}
Dasmeh,~P.; Searles,~D.~J.; Ajloo,~D.; Evans,~D.~J.; Williams,~S.~R. {On
  violations of Le Chatelier’s principle for a temperature change in small
  systems observed for short times}. \emph{The Journal of Chemical Physics}
  \textbf{2009}, \emph{131}, 214503\relax
\mciteBstWouldAddEndPuncttrue
\mciteSetBstMidEndSepPunct{\mcitedefaultmidpunct}
{\mcitedefaultendpunct}{\mcitedefaultseppunct}\relax
\EndOfBibitem
\bibitem[Dasmeh \latin{et~al.}(2011)Dasmeh, Ajloo, and Searles]{dasmeh2011}
Dasmeh,~P.; Ajloo,~D.; Searles,~D.~J. Transient violation of Le Chatelier’s
  principle for a network of water molecules. \emph{Journal of the Iranian
  Chemical Society} \textbf{2011}, \emph{8}, 424--432\relax
\mciteBstWouldAddEndPuncttrue
\mciteSetBstMidEndSepPunct{\mcitedefaultmidpunct}
{\mcitedefaultendpunct}{\mcitedefaultseppunct}\relax
\EndOfBibitem
\bibitem[Evans \latin{et~al.}(1993)Evans, Cohen, and Morriss]{Evans1993}
Evans,~D.~J.; Cohen,~E. G.~D.; Morriss,~G.~P. {Probability of second law
  violations in shearing steady states}. \emph{Physical Review Letters}
  \textbf{1993}, \emph{71}, 2401--2404\relax
\mciteBstWouldAddEndPuncttrue
\mciteSetBstMidEndSepPunct{\mcitedefaultmidpunct}
{\mcitedefaultendpunct}{\mcitedefaultseppunct}\relax
\EndOfBibitem
\bibitem[Evans and Searles(1994)Evans, and Searles]{Evans1994}
Evans,~D.~J.; Searles,~D.~J. Equilibrium microstates which generate second law
  violating steady states. \emph{Physical Review E} \textbf{1994}, \emph{50},
  1645--1648\relax
\mciteBstWouldAddEndPuncttrue
\mciteSetBstMidEndSepPunct{\mcitedefaultmidpunct}
{\mcitedefaultendpunct}{\mcitedefaultseppunct}\relax
\EndOfBibitem
\bibitem[Evans and Searles(1996)Evans, and Searles]{Evans1996}
Evans,~D.~J.; Searles,~D.~J. Causality, response theory, and the second law of
  thermodynamics. \emph{Physical Review E} \textbf{1996}, \emph{53},
  5808--5815\relax
\mciteBstWouldAddEndPuncttrue
\mciteSetBstMidEndSepPunct{\mcitedefaultmidpunct}
{\mcitedefaultendpunct}{\mcitedefaultseppunct}\relax
\EndOfBibitem
\bibitem[Evans \latin{et~al.}(2012)Evans, Williams, and
  Rondoni]{evans2012mathematical}
Evans,~D.~J.; Williams,~S.~R.; Rondoni,~L. A mathematical proof of the zeroth
  “law” of thermodynamics and the nonlinear Fourier “law” for heat
  flow. \emph{The Journal of Physical Chemistry} \textbf{2012}, \emph{137},
  194109\relax
\mciteBstWouldAddEndPuncttrue
\mciteSetBstMidEndSepPunct{\mcitedefaultmidpunct}
{\mcitedefaultendpunct}{\mcitedefaultseppunct}\relax
\EndOfBibitem
\bibitem[Jarzynski(2011)]{Jarzynski2011}
Jarzynski,~C. {Equalities and inequalities: Irreversibility and the Second Law
  of Thermodynamics at the nanoscale}. \emph{Annual Review of Condensed Matter
  Physics} \textbf{2011}, \emph{2}, 329--351\relax
\mciteBstWouldAddEndPuncttrue
\mciteSetBstMidEndSepPunct{\mcitedefaultmidpunct}
{\mcitedefaultendpunct}{\mcitedefaultseppunct}\relax
\EndOfBibitem
\bibitem[Carberry \latin{et~al.}(2004)Carberry, Reid, Wang, Sevick, Searles,
  and Evans]{Carberry2004140601}
Carberry,~D.; Reid,~J.; Wang,~G.; Sevick,~E.; Searles,~D.; Evans,~D.
  Fluctuations and irreversibility: An experimental demonstration of a
  second-law-like theorem using a colloidal particle held in an optical trap.
  \emph{Physical Review Letters} \textbf{2004}, \emph{92}\relax
\mciteBstWouldAddEndPuncttrue
\mciteSetBstMidEndSepPunct{\mcitedefaultmidpunct}
{\mcitedefaultendpunct}{\mcitedefaultseppunct}\relax
\EndOfBibitem
\bibitem[Wang \latin{et~al.}(2002)Wang, Sevick, Mittag, Searles, and
  Evans]{Wang2002PRL}
Wang,~G.~M.; Sevick,~E.~M.; Mittag,~E.; Searles,~D.~J.; Evans,~D.~J.
  {Experimental demonstration of violations of the Second Law of Thermodynamics
  for small systems and short time scales}. \emph{Physical Review Letters}
  \textbf{2002}, \emph{89}, 050601\relax
\mciteBstWouldAddEndPuncttrue
\mciteSetBstMidEndSepPunct{\mcitedefaultmidpunct}
{\mcitedefaultendpunct}{\mcitedefaultseppunct}\relax
\EndOfBibitem
\bibitem[Evans \latin{et~al.}(2011)Evans, Williams, and Searles]{Evans2011}
Evans,~D.; Williams,~S.; Searles,~D. A proof of Clausius' theorem for time
  reversible deterministic microscopic dynamics. \emph{Journal of Chemical
  Physics} \textbf{2011}, \emph{134}\relax
\mciteBstWouldAddEndPuncttrue
\mciteSetBstMidEndSepPunct{\mcitedefaultmidpunct}
{\mcitedefaultendpunct}{\mcitedefaultseppunct}\relax
\EndOfBibitem
\bibitem[Evans \latin{et~al.}(2016)Evans, Searles, and
  Williams]{Evans2016aderivation}
Evans,~D.; Searles,~D.; Williams,~S. A derivation of the Gibbs equation and the
  determination of change in Gibbs entropy from calorimetry. \emph{Australian
  Journal of Chemistry} \textbf{2016}, \emph{69}, 1413--1419\relax
\mciteBstWouldAddEndPuncttrue
\mciteSetBstMidEndSepPunct{\mcitedefaultmidpunct}
{\mcitedefaultendpunct}{\mcitedefaultseppunct}\relax
\EndOfBibitem
\bibitem[Evans \latin{et~al.}(2010)Evans, Searles, and
  Williams]{Evans2010Fourier}
Evans,~D.; Searles,~D.; Williams,~S. On the probability of violations of
  Fourier's law for heat flow in small systems observed for short times.
  \emph{Journal of Chemical Physics} \textbf{2010}, \emph{132}\relax
\mciteBstWouldAddEndPuncttrue
\mciteSetBstMidEndSepPunct{\mcitedefaultmidpunct}
{\mcitedefaultendpunct}{\mcitedefaultseppunct}\relax
\EndOfBibitem
\bibitem[Onuki(2019)]{Onuki2019}
Onuki,~A. {Theory of applying heat flow from thermostatted boundary walls:
  Dissipative and local-equilibrium responses and fluctuation theorems}.
  \emph{The Journal of Chemical Physics} \textbf{2019}, \emph{151},
  134118\relax
\mciteBstWouldAddEndPuncttrue
\mciteSetBstMidEndSepPunct{\mcitedefaultmidpunct}
{\mcitedefaultendpunct}{\mcitedefaultseppunct}\relax
\EndOfBibitem
\bibitem[Petersen \latin{et~al.}(2016)Petersen, Evans, and
  Williams]{petersen2016dissipation}
Petersen,~C.~F.; Evans,~D.~J.; Williams,~S.~R. Dissipation in monotonic and
  non-monotonic relaxation to equilibrium. \emph{The Journal of Chemical
  Physics} \textbf{2016}, \emph{144}, 074107\relax
\mciteBstWouldAddEndPuncttrue
\mciteSetBstMidEndSepPunct{\mcitedefaultmidpunct}
{\mcitedefaultendpunct}{\mcitedefaultseppunct}\relax
\EndOfBibitem
\bibitem[Petersen \latin{et~al.}(2016)Petersen, Evans, and
  Williams]{petersen2016mechanism}
Petersen,~C.~F.; Evans,~D.~J.; Williams,~S.~R. Mechanism for asymmetric bias in
  demonstrations of the NPI and fluctuation theorem. \emph{Molecular
  Simulation} \textbf{2016}, \emph{42}, 531--541\relax
\mciteBstWouldAddEndPuncttrue
\mciteSetBstMidEndSepPunct{\mcitedefaultmidpunct}
{\mcitedefaultendpunct}{\mcitedefaultseppunct}\relax
\EndOfBibitem
\bibitem[Petersen and Searles(2022)Petersen, and Searles]{petersen2022}
Petersen,~C.~F.; Searles,~D.~J. Equilibrium distribution functions: connection
  with microscopic dynamics. \emph{Physical Chemistry Chemical Physics}
  \textbf{2022}, \emph{24}, 6383--6392\relax
\mciteBstWouldAddEndPuncttrue
\mciteSetBstMidEndSepPunct{\mcitedefaultmidpunct}
{\mcitedefaultendpunct}{\mcitedefaultseppunct}\relax
\EndOfBibitem
\bibitem[Crooks(1999)]{Crooks1999}
Crooks,~G.~E. {Entropy production ﬂuctuation theorem and the nonequilibrium
  work relation}. \emph{Physical Review E} \textbf{1999}, \emph{60},
  2721--2726\relax
\mciteBstWouldAddEndPuncttrue
\mciteSetBstMidEndSepPunct{\mcitedefaultmidpunct}
{\mcitedefaultendpunct}{\mcitedefaultseppunct}\relax
\EndOfBibitem
\bibitem[Petersen and Searles(2022)Petersen, and Searles]{Petersen20226383}
Petersen,~C.; Searles,~D. Equilibrium distribution functions: connection with
  microscopic dynamics. \emph{Physical Chemistry Chemical Physics}
  \textbf{2022}, \emph{24}, 6383--6392\relax
\mciteBstWouldAddEndPuncttrue
\mciteSetBstMidEndSepPunct{\mcitedefaultmidpunct}
{\mcitedefaultendpunct}{\mcitedefaultseppunct}\relax
\EndOfBibitem
\bibitem[Searles and Evans(2000)Searles, and Evans]{Searles20003503}
Searles,~D.; Evans,~D. Ensemble dependence of the transient fluctuation
  theorem. \emph{Journal of Chemical Physics} \textbf{2000}, \emph{113},
  3503--3509\relax
\mciteBstWouldAddEndPuncttrue
\mciteSetBstMidEndSepPunct{\mcitedefaultmidpunct}
{\mcitedefaultendpunct}{\mcitedefaultseppunct}\relax
\EndOfBibitem
\bibitem[Thompson \latin{et~al.}(2022)Thompson, Aktulga, Berger, Bolintineanu,
  Brown, Crozier, in~'t Veld, Kohlmeyer, Moore, Nguyen, Shan, Stevens,
  Tranchida, Trott, and Plimpton]{LAMMPS}
Thompson,~A.~P.; Aktulga,~H.~M.; Berger,~R.; Bolintineanu,~D.~S.; Brown,~W.~M.;
  Crozier,~P.~S.; in~'t Veld,~P.~J.; Kohlmeyer,~A.; Moore,~S.~G.;
  Nguyen,~T.~D.; Shan,~R.; Stevens,~M.~J.; Tranchida,~J.; Trott,~C.;
  Plimpton,~S.~J. {LAMMPS} - {A} flexible simulation tool for particle-based
  materials modeling at the atomic, meso, and continuum scales. \emph{Computer
  Physics Communications} \textbf{2022}, \emph{271}, 108171\relax
\mciteBstWouldAddEndPuncttrue
\mciteSetBstMidEndSepPunct{\mcitedefaultmidpunct}
{\mcitedefaultendpunct}{\mcitedefaultseppunct}\relax
\EndOfBibitem
\end{mcitethebibliography}
\end{document}